\documentclass[12pt]{article}
\usepackage[centertags]{amsmath}
\usepackage{amsmath, amsthm, amssymb, amsfonts, amscd, xspace, pifont, natbib, subfigure, changepage, graphicx, setspace, lineno}
\usepackage{epsfig, amsfonts, verbatim, multirow, hyperref, graphicx, rotating, booktabs, bbm}
\usepackage{url, algorithm, color, xcolor}

\newcommand{\mbf}{\mathbf}
\newcommand{\bs}{\boldsymbol}
\newtheorem{thm}{Theorem}[section]

\newtheorem{prop}[thm]{Proposition}
\theoremstyle{definition}
\newtheorem{defn}[thm]{Definition}
\theoremstyle{remark}

\def\boxit#1{\vbox{\hrule\hbox{\vrule\kern6pt
          \vbox{\kern6pt#1\kern6pt}\kern6pt\vrule}\hrule}}

\newcommand{\pal}{\partial}

\newcommand{\ba}{\mbox{\bf a}}
\newcommand{\bb}{\mbox{\bf b}}

\newcommand{\bh}{\mbox{\bf h}}

\newcommand{\bu}{\mbox{\bf u}}
\newcommand{\bv}{\mbox{\bf v}}
\newcommand{\bw}{\mbox{\bf w}}

\newcommand{\bA}{\mbox{\bf A}}
\newcommand{\bB}{\mbox{\bf B}}

\newcommand{\bD}{\mbox{\bf D}}

\newcommand{\bH}{\mbox{\bf H}}

\newcommand{\bI}{\mbox{\bf I}}

\newcommand{\bQ}{\mbox{\bf Q}}

\newcommand{\bS}{\mbox{\bf S}}
\newcommand{\bU}{\mbox{\bf U}}

\newcommand{\bV}{\mbox{\bf V}}

\newcommand{\bX}{\mbox{\bf X}}

\newcommand{\bY}{\mbox{\bf Y}}
\newcommand{\bZ}{\mbox{\bf Z}}

\newcommand{\what}{\widehat}

\newcommand{\diag}{\mathrm{diag}}

\def\t{^\top}

\def\beqn{\begin{eqnarray}}
\def\eeqn{\end{eqnarray}}
\def\beqns{\begin{eqnarray*}}
\def\eeqns{\end{eqnarray*}}

\begin{document}
\pagewiselinenumbers
\setlength\linenumbersep{20pt}
\renewcommand\linenumberfont{\normalfont\tiny\sffamily\color{gray}}
\begin{titlepage}
\title{\LARGE On The Degrees of Freedom of Reduced-rank Estimators in Multivariate Regression}
\author{Ashin Mukherjee$^*$, Kun Chen$^+$, Naisyin Wang$^*$ and Ji Zhu$^*$\\
$^*$University of Michigan, $^+$Kansas State University}
\maketitle
\vspace{0.3in}

\abstract{\noindent In this paper we study the effective degrees of freedom of a general class of reduced rank estimators for multivariate regression in the framework of Stein's unbiased risk estimation (SURE). We derive a finite-sample exact unbiased estimator that admits a closed-form expression in terms of the singular values or thresholded singular values of the least squares solution and hence readily computable. The results continue to hold in the high-dimensional scenario when both the predictor and response dimensions are allowed to be larger than the sample size. The derived analytical form facilitates the investigation of its theoretical properties and provides new insights into the empirical behaviors of the degrees of freedom. In particular, we examine the differences and connections between the proposed estimator and a commonly-used naive estimator, i.e., the number of free parameters. The use of the proposed estimator leads to efficient and accurate prediction risk estimation and model selection, as demonstrated by simulation studies and a data example.}


\thispagestyle{empty}
\end{titlepage}

\setcounter{equation}{0}

\section{Introduction}
Multivariate linear regression is the extension of the classical univariate regression model to the case where we have $q(>1)$ responses and $p$ predictors. It is commonly used in bioinformatics, chemometrics, econometrics, and other quantitative fields where one is interested in predicting several responses simultaneously.\\

\noindent We can express the multivariate linear regression model in matrix notation as follows. Let $\mbf{X}$ denote the $n \times p$ predictor or design matrix, with the $i$-th row $\mbf{x}_i =  \left(x_{i1},x_{i2},\ldots, x_{ip}\right) \in \mathbb{R}^p$. Similarly the $n \times q$ dimensional response matrix is denoted by $\mbf{Y}$, where the $i$-th row is $\mbf{y}_i = \left(y_{i1},y_{i2},\ldots,y_{iq}\right) \in \mathbb{R}^q$. The regression parameters are given by the coefficient matrix $\mbf{B}$ which is of dimension $p \times q$. Note that the $k$-th column of $\mbf{B}$ is the regression coefficient vector for regressing the $k$-th response on the predictors. Let $\mbf{E}$ denote the $n \times q$ random error matrix with independent entries with mean zero and variance $\sigma^2$. Then the multivariate linear regression model is given by
\begin{equation}\label{model}
\rm \mbf{Y} = \mbf{XB} + \mbf{E}.
\end{equation}

\noindent Note that, this reduces to the classical univariate regression model when $q=1$. For notational simplicity, we assume that the responses and the predictors are centered, and hence the intercept term can be omitted without any loss of generality. The ordinary least squares approach of estimating $\mbf{B}$ leads to
\[
\widehat{\mbf{B}}_{\mbox{ols}} = \left(\mbf{X}\t \mbf{X}\right)^{-1}\mbf{X}\t \mbf{Y}.
\]
The ordinary least squares estimate amounts to performing $q$ separate univariate regressions and completely ignores the multivariate aspect of the problem, where many of the responses might be highly correlated and hence the effective dimensionality can be much smaller than $q$. Also it is unsuitable for the high-dimensional case where both $p, q > n$. Quite a large number of methods have been proposed in the literature to overcome these drawbacks.  Many of them would fall under the general class of \textit{linear factor regression}, where the responses are regressed against a small number of linear combination of predictors commonly known as factors.  Examples include principal component regression \citep{Massy1965}, partial least squares \citep{Wold1975}, canonical correlation analysis \citep{Hotelling1935} and so on.  The methods differ in the way they choose the factors. Recently \citet{Witten2009} introduced a penalized canonical correlation analysis using sparse matrix factorization that leads to more interpretable factors and is more suitable for high-dimensional problems. \citet{Breiman1997} proposed the curds and whey(C\&W) approach which borrows strength by performing a second round of regression of the responses on the  ordinary least squares  estimators.  The authors also show some close connections of the C\&W approach with canonical correlation analysis. \\

\noindent Yet another line of research focuses on the rank of the regression coefficient matrix. \citet{Anderson1951} proposed a class of regression models that restrict the rank of the coefficient matrix to be much smaller than the dimensionality of $\mbf{B}$, i.e. $rank(\mbf{B}) \leq r \leq \min\{p, q\}$.  This is a quite reasonable assumption in many multivariate regression problems, which can be interpreted as follows: the $q$ responses are related to the $p$ predictors only through $r$ effective linear factors.  It results in the following optimization problem
\begin{equation}\label{RRR}
\widehat{\mbf{B}}(r) = \underset{\{\mbf{B} : \textrm{rank}(\mbf{B}) \leq r\}}{\rm argmin} \lVert \mbf{Y} - \mbf{XB} \rVert^2_F,
\end{equation}

\noindent where $\lVert . \rVert_F$ denotes the Frobenius norm of a matrix. Even though the rank penalty makes it a non-convex optimization problem, it admits a closed form solution as we shall see later. \citet{Izenman1975} introduced the term reduced rank regression for this class of models and derived the asymptotic distributions and confidence intervals for reduced rank regression estimators. A non-exhaustive  list of notable work includes \citet{Rao1978}, \citet{Davies1982}, \citet{Anderson1999, Anderson2002}; see \citet{Reinsel1998} or \citet{Izenman2008} for a more comprehensive account. Recently, there has been a revival of interest in the reduced rank methods. Instead of restricting the rank, \citet{Yuan2007} proposed to put an $\ell_1$ penalty on the singular values of $\mbf{B}$ also known as the nuclear norm. The nuclear-norm penalized least squares criterion encourages sparsity among the singular values to achieve simultaneous rank reduction and shrinkage coefficient estimation \citep{Neghaban2011, Lu2012}. However, this method is computationally intensive and tends to overestimate the rank \citep{Bunea2011}. \citet{Bunea2012} proposed the rank selection criterion extending reduced rank regression to high-dimensional settings, in which rank-constrained estimation was cast as a penalized least squares method with the penalty proportional to the rank of the coefficient matrix, or equivalently, the $\ell_0$ norm of its singular values. Under that framework the authors were able to characterize the choice of tuning parameter, which guarantees asymptotic consistency in terms of rank selection. \citet{Chen2012JRSSB} adopted sparsity penalties on singular vectors for reduced rank regression problems that lead to more interpretable models. Very recently \citet{Chen2012} proposed an adaptive nuclear norm penalty on the signal matrix $\mbf{XB}$ aiming to close the gap between $\ell_0$ and $\ell_1$ penalties on singular values. The resulting optimization problem admits a closed form solution and enjoys many desirable theoretical properties. \\

\noindent In this paper we study the degrees of freedom of the reduced rank estimators in multivariate linear regression models. The degrees of freedom is a very familiar and one of the most widely used terms in statistics.  We utilize it from ANOVA t-tests to model selection criteria such as AIC and BIC. However, it has been largely overlooked in the reduced rank regression literature except for some heuristic suggestions \citep{Davies1982, Reinsel1998}. For example, the number of free parameters in a $p \times q$ matrix of rank $r$, given by $r(p + q - r)$ has been suggested as a naive estimate of the degrees of freedom of the reduced rank regression estimator when restricted to rank $r \leq \min\{p, q\}$. More precisely, for an arbitrary design matrix, the number of free parameters should be $(r_x+q-r)r$, where $r_x=\mbox{rank}(\bX)$ is the rank of the design matrix \citep{Bunea2011}. Henceforth, we refer to this as the naive estimator of the degrees of freedom of a rank-$r$ model. In this paper, we aim to find a finite-sample unbiased estimator of the degrees of freedom for a general class of reduced rank estimators for the multivariate regression model and investigate its properties. The result covers a significant gap in the literature, as the previously suggested naive estimate lacks both statistical motivation and practical performance. \\

\noindent In a nutshell, the degrees of freedom quantifies the complexity of a statistical modeling procedure \citep{Hastie1990}. In the case of the univariate linear regression model, it is well-known that the degrees of freedom is the number of estimated parameters, $p$.  However, in general there is no exact correspondence between the degrees of freedom and the number of free parameters in the model \citep{Ye1998}.  For example, in the best subset selection for univariate regression \citep{Hocking1967}, we search for the best model of size $p_0 \in \{1, 2, \ldots, p\}$ that minimizes the residual sum of squares.  The resulting model has $p_0$ parameters but intuitively the degrees of freedom would be higher than $p_0$ since the search for the ``optimal'' subset of size $p_0$ increases model complexity \citep{Hastie2009}.  In other words, for best subset selection the optimal $p_0$-dimensional subspace that minimizes the residual sum of squares clearly depends on $\mbf{y}$.  Thus the final estimator is highly non-linear in $\mbf{y}$, which results in the loss of correspondence between degrees of freedom and the number of parameters in the model.\\

\noindent Similar arguments also apply to the reduced rank regression.  Instead of searching for best $p_0$-variables as in the case of best subset selection, here we are searching for best $r$ linear combinations of the predictors that minimize the least squares loss, which should intuitively suggest increased model complexity.  Since the optimal rank $r$-subspace depends on the response matrix $\mbf{Y}$, the natural correspondence between number of free parameters and degrees of freedom need not hold.  This is where reduced rank regression is different from other linear factor regression methods, e.g. principal component regression \citep{Massy1965}. In principal component regression, the factors are principal components of the design matrix $\mbf{X}$, which do not depend on the response $\mbf{Y}$, thus the final estimator is still linear in $\mbf{Y}$. \\

\noindent The rest of the paper is organized as follows.  In section \ref{sec:Intro}, we review the degrees of freedom in the framework of Stein's unbiased risk estimation \citep{Stein1981}. The reduced rank regression estimator is discussed in detail in section \ref{sec:reduced rank regressionclass}, additionally, we also introduce a more general class of reduced rank estimators. Sections \ref{sec:Deriv}, \ref{sec:ClosedForm} and \ref{sec:Existence} contain the main results on our proposed exact unbiased estimator of  the degrees of freedom including derivation of a closed form expression, connections to naive degrees of freedom and almost everywhere existence. In section \ref{sec:Sim}, we show that the exact unbiased estimator of the degrees of freedom for reduced rank regression methods can be significantly different from the naive estimator through several numerical examples.  We also show that using the exact unbiased estimate of degrees of freedom can lead to gain in prediction accuracy over its heuristic counterpart. In section \ref{sec:Dat}, we apply the developed method to a genetic association study, and we conclude the paper with a discussion in section \ref{sec:Sum}.

\section{Degrees of freedom}\label{sec:Intro}

\noindent \citet{Stein1981} in his theory of unbiased risk estimation (SURE) first introduced a rigorous definition of the degrees of freedom of a statistical estimation procedure.  Later \citet{Efron2004} showed that Stein's treatment can be considered as a special case of a more general notion under the assumption of Gaussianity.  Assume that we have data of the form $\left(\mbf{y}_{n \times 1}, \mbf{X}_{n \times p}\right)$.  Given $\mbf{X}$, the response originates from the following model $\mbf{y} \sim (\bs{\mu}, \sigma^2 \mbf{I})$, where $\bs{\mu}$ is the true mean that can be a function of $\mbf{X}$, and $\sigma^2$ is the common variance. Then for any estimation procedure $m(\cdot)$ with fitted values $\hat{\bs{\mu}} = m(\mbf{X}, \mbf{y})$, the degrees of freedom of $m(\cdot)$ is defined as
\begin{equation}\label{dfOrig}
\displaystyle df(m) = \sum_{i = 1}^n cov(\hat{\mu}_i, y_i)/\sigma^2.
\end{equation}

\noindent The rationale is that more complex models would try to fit the data better, and hence the covariance between observed and fitted pairs would be higher.  This expression is not directly observable except for certain simple cases, for example, when $m(\mbf{y}) = \mbf{Sy}$, a linear smoother.  In that case, it is not difficult to see that $df(m) = tr(\mbf{S})$.  Stein was able to overcome this hurdle for a special case when $\mbf{y} \sim N(\bs{\mu}, \sigma^2 \mbf{I})$.  Using a simple equality for the Gaussian distribution, he proved that as long as the partial derivative $\partial \hat{\mu}_i / \partial y_i$ exists almost everywhere for all $i \in \{1,2,\ldots,n\}$, the following holds
\[
cov(\hat{\mu}_i, y_i) = \sigma^2 \mathbb{E}\left(\frac{\partial \hat{\mu}_i}{\partial y_i}\right).
\]

\noindent Thus, we have the following unbiased estimator of the degrees of freedom for the fitting procedure $m(\cdot)$
\begin{equation}\label{dfUE}
\displaystyle \widehat{df} = \sum_{i = 1}^n \frac{\partial \hat{\mu}_i}{\partial y_i}.
\end{equation}

\noindent Using the degrees of freedom definition as in \eqref{dfOrig}, \citet{Efron2004} employed the covariance penalty approach to prove that the $C_p$-type statistics \citep{Mallow1973} is an unbiased estimator of the true prediction error, where
\begin{equation}\label{Cp}
\displaystyle C_p(\hat{\bs{\mu}}) = \frac{1}{n}\lVert \mbf{y} - \hat{\bs{\mu}} \rVert^2 + \frac{2df(\hat{\bs{\mu}})}{n}\sigma^2.
\end{equation}

\noindent This reveals the important role played by the degrees of freedom in model assessment.  It gives us a principled way of selecting the optimal model without going for computationally expensive methods such as cross-validation, and in certain settings it can offer significantly better prediction accuracy than such methods \citep{Efron2004}.  
Indeed the degrees of freedom is an integral part of almost every model selection criterion, including Bayesian Information Criterion (BIC) \citep{Schwarz1978}, generalized cross-validation (GCV) \citep{Golub1979} and so on. Many important works followed that of \citet{Stein1981} and \citet{Efron2004}.  For example, \citet{Donoho1995} used the SURE theory to derive the degrees of freedom for the soft-thresholding operator in wavelet shrinkage; \citet{Meyer2000} employed this framework to derive the same for shape restricted regression; \citet{Li2008} also used this set-up to derive an unbiased estimator of the degrees of freedom for penalized quantile regression.  \citet{Zou2007} applied the SURE theory for the popular regression shrinkage and variable selection method lasso \citep{Tibshirani1996}.  This is a challenging problem because of the non-linear nature of lasso solution, which does not admit an analytical solution except for certain special cases.  Using sophisticated mathematical analysis, \citet{Zou2007} were able to show that the number of non-zero coefficients provides an unbiased estimate of the degrees of freedom for the lasso.  This is a result of great practical importance since this allows one to come up with model selection criteria such as $C_p$ and BIC for the lasso without incurring any extra computational cost. \\

\noindent The degrees of freedom for the reduced rank estimators also proves to be a challenging problem because of the non-linearity of the estimator.  As we will see shortly, even though it admits a closed-form solution, the solution is highly non-linear depending on singular value decomposition of the least squares solution $\widehat{\mbf{Y}}$ described in \eqref{ols}.  In the next several sections, we study the degrees of freedom of a general class of reduced rank estimators in the framework of SURE and propose a finite-sample exactly unbiased estimator. The importance of such an estimator has been emphasized repeatedly by \citet{Shen2002}, \citet{Efron2004}, \citet{Zou2007} among others. \\

\noindent To overcome the analytical difficulty in computing the degrees of freedom, \citet{Ye1998} and \citet{Shen2002} proposed the generalized degrees of freedom approach, where they evaluate \eqref{dfUE} numerically, using data perturbation techniques to compute an approximately unbiased estimator of the degrees of freedom.  \citet{Efron2004} also proposed a bootstrap based idea to arrive at an approximately unbiased estimator of \eqref{dfOrig}. Though these kind of simulation based approaches allow us to extend the degrees of freedom approach to many highly non-linear modeling frameworks, they are computationally expensive. Also this type of numerical solutions does not admit any closed-form expression making investigation of the theoretical properties an extremely difficult task, thus limiting our insight.

\section{A class of reduced rank estimators}\label{sec:reduced rank regressionclass}

\noindent  
Recall the multivariate linear regression model as in \eqref{model}. Let $\widehat{\mbf{Y}}$ be the least squares estimate which admits a singular value decomposition of the form
\begin{equation}\label{ols}
\widehat{\mbf{Y}} = \mbf{X}(\mbf{X}\t \mbf{X})^{+} \mbf{X}\t \mbf{Y} = \underset{n \times \bar{r}}{\mbf{W}}  \ \underset{\bar{r} \times \bar{r}}{\mbf{D}} \ \underset{\bar{r} \times q}{\mbf{V}\t},
\end{equation}
where $(\mbf{A})^+$ denotes the Moore-Penrose inverse \citep{Moore1920, Penrose1955} of a generic matrix $\mbf{A}$. Note that this is well defined even when $p, q > n$ or the design matrix $\mbf{X}$ is of low rank. $\mbf{W}$ and $\mbf{V}$ are orthogonal matrices that represent the left and right singular vectors and $\mbf{D} = diag\{d_i, i = 1,\ldots,\bar{r}\}$ with $d_1\geq \cdots \geq d_{\bar{r}} > 0$ are the singular values of $\widehat{\mbf{Y}}$.  Without loss of generality we assume that, $rank(\widehat{\mbf{Y}}) = \bar{r} = \min (r_x, q)$, where $r_x$ denotes the rank of the design matrix. We will denote the $k$-th column of $\mbf{W}$ and $\mbf{V}$ by $w_k$ and $v_k$ respectively.  Using the Eckart-Young theorem \citep{Eckart1936}, it is not difficult to show that the reduced rank regression estimator for \eqref{RRR} can be expressed as
\begin{equation}\label{RRRsol}
\widehat{\mbf{Y}}(r) = \widehat{\mbf{Y}}\sum_{k=1}^r v_k v_k\t = \mbf{W}^{(r)} \mbf{D}^{(r)} \mbf{V}^{(r)\top}, \quad r = 1, \ldots, \bar{r}, 
\end{equation} 
where $\mbf{A}^{(r)}$ denotes the first $r$-columns of a generic matrix $\mbf{A}$. This rank constrained estimation procedure can also be viewed under a more general penalized least squares framework
\begin{equation}\label{pc1}
\min_{\mbf{B}}\left\{ \frac{1}{2} \lVert \mbf{Y} - \mbf{XB}\rVert^2_F + \lambda \mathcal{P}(\mbf{B})\right\},
\end{equation}
in which the penalty is proportional to the rank of the coefficient matrix $\mbf{B}$, i.e., $\mathcal{P}(\mbf{B}) = rank(\mbf{B})$ \citep{Bunea2011}, and it leads to a hard-thresholding of the singular values of $\widehat{\mbf{Y}}$. More generally, under the regularized estimation framework \eqref{pc1}, a set of reduced-rank estimators may be indexed by the regularization parameter $\lambda$, which controls the penalty level and hence the model's complexity. In light of that, we consider a broad class of such reduced-rank estimators defined as
\begin{equation}
\tilde{\bY}(\lambda) = \bX\tilde{\bB}(\lambda)= \sum_{k=1}^{\bar{r}}s_k(d_k,\lambda)d_k\bw_k\bv_k\t=\what{\bY}\sum_{k=1}^{\bar{r}}s_k(d_k,\lambda)\bv_k\bv_k\t,\label{ytilde}
\end{equation}
\noindent where each $s_k(d_k,\lambda) \in [0,1]$ is a function of $d_k$ and $\lambda$, and they satisfy $s_1(d_1,\lambda)\geq \cdots\geq s_{\bar{r}}(d_{\bar{r}},\lambda)\geq 0$. To avoid confusion, we may simply write $s_k(d_k,\lambda) = s_k(\lambda) = s_k$. The reduced rank regression estimator can be viewed as a special case of this general framework with $s_k(d_k, r) = \mathbbm{1}(k \leq r) \in \{0,1\}$, $r = 1, \ldots, \bar{r}$, where the solutions are indexed by the rank constraint $r$, instead of a continuous penalty parameter $\lambda$. Note that this class of estimators has the same set of singular vectors as the reduced rank regression estimator in \eqref{RRRsol}, but may have different singular value estimates given by some shrunk or thresholded versions of the estimated singular values from least squares. Such estimators can be obtained from a non-convex singular-value penalization or thresholding operations \citep{She2009, She2012, Chen2012}. The class of estimators \eqref{ytilde} is computationally efficient and possesses many desirable theoretical properties, such as, rank selection consistency and achieving minimax error bound \citep{Bunea2011, Chen2012} under both the classical and the high-dimensional asymptotic regimes. Some examples include the reduced rank regression, rank selection criterion \citep{Bunea2011}, the nuclear norm penalized estimator under an orthogonal design \citep{Yuan2007}, and the adaptive nuclear norm estimator proposed by \citet{Chen2012}. 



\section{Degrees of freedom of reduced rank estimators}\label{sec:Deriv}

In the previous section we discussed a broad class of reduced rank estimators covering both hard-thresholding and soft-thresholding of the singular values of $\widehat{\mbf{Y}}$. Next we apply definition \eqref{dfUE} to such multivariate regression estimators to estimate the degrees of freedom. To answer that we start by rewriting the multivariate linear regression model \eqref{model} as follows 
\[
\displaystyle \underset{nq \times 1}{vec(\mbf{Y})} = \underset{nq \times pq}{\left[\mbf{I}_q \otimes \mbf{X}\right]} \underset{pq \times 1}{vec(\mbf{B})} + \underset{nq \times 1}{vec(\mbf{E})},
\]

\noindent where $\otimes$ denotes the usual Kronecker product between matrices, and $vec(\cdot)$ stands for the column-wise vectorization operator on a matrix. We will first derive the results for the special case of reduced rank regression estimator \eqref{RRRsol} and later extend it to the general class of model \eqref{ytilde}. Applying definition \eqref{dfUE} we get

\begin{equation}\label{est}
\widehat{df}(r) = tr \left\{ \frac{\partial vec(\widehat{\mbf{Y}}(r))}{\partial vec(\mbf{Y})}\right\}, \quad r = 1,\ldots,\bar{r},
\end{equation}

\noindent where $tr(\cdot)$ denotes the trace operator for a real square matrix. Recall that we assumed $rank(\widehat{\mbf{Y}}) = \bar{r} = \min(r_x, q)$ which is not restrictive in general and does not depend on the dimensions of the problem. Let $\bX\t\bX=\bQ\bS^2\bQ\t$ be the eigen decomposition of $\bX\t\bX$, i.e., $\bQ\in \mathbb{R}^{p\times r_x}$, $\bQ\t\bQ=\bI$, and $\bS\in\mathbb{R}^{r_x\times r_x}$ is a diagonal matrix with positive diagonal elements. Then, the Moore-Penrose inverse of $\bX\t\bX$ can be written as $(\bX\t\bX)^+=\bQ\bS^{-2}\bQ\t$. Define

\begin{equation*}
\bH=\bS^{-1}\bQ\t\bX\t\bY.
\end{equation*}
It then follows that $\bH\in\mathbb{R}^{r_x\times q}$ admits an SVD of the form
\begin{equation}
\bH=\bU\bD\bV\t,\label{svdh}
\end{equation}
where $\bU\in \mathbb{R}^{r_x\times \bar{r}}$, $\bU\t\bU=\bI$, and $\bV$, $\bD$ are defined in \eqref{ols}. The matrix $\bH$ shares the same set of singular values and right singular vectors with $\what{\bY}$ in \eqref{ols}, as $\bH\t\bH = \what{\bY}\t \what{\bY} = \bY\t \bX (\bX\t\bX)^+ \bX\t \bY$. Moreover, $\bH$ is full rank since $\what{\bY}$ is of rank $\bar{r} = \min(r_x,q)$. The matrix $\bH$ plays a key role in deriving a simple form of the degrees of freedom as we shall see later. In particular, this construction allows us to avoid singularities arising from $r_x < p$ in the high-dimensional scenario. Simplifying \eqref{est} using matrix equalities such as $tr(\mbf{AB}) = tr(\mbf{BA})$ and $vec(\mbf{ABC}) = (\mbf{C}\t \otimes \mbf{A})vec(\mbf{B})$ we obtain our unbiased estimator of the degrees of freedom of reduced rank regression as

\begin{equation}
\what{df}(r)=tr\left\{\frac{\partial \mbox{vec}(\bU^{(r)}\bD^{(r)}\bV^{(r)\top})}{\partial \mbox{vec}(\bH)}\right\} = tr\left\{\frac{\partial \mbox{vec}(\bH(r))}{\partial \mbox{vec}(\bH)}\right\}=\sum_{i=1}^{r_x}\sum_{j=1}^{q}\frac{\partial h_{ij}(r)}{\partial h_{ij}},\label{df}
\end{equation}
where $\bH(r)=\bU^{(r)}\bD^{(r)}\bV^{(r)\top}=(h_{ij}(r))_{r_x\times q}$ is the rank $r$ approximation to $\mbf{H}$. The details of this derivation could be found in the Appendix. For the general class of reduced-rank estimators in \eqref{ytilde}, we have

\begin{align*}
\tilde{\bY}(\lambda) = \bX\bQ\bS^{-1}\bH\sum_{k=1}^{\bar{r}}s_k(d_k,\lambda) \bv_k\bv_k\t = \bX\bQ\bS^{-1}\bU\tilde{\bD}(\lambda)\bV^{\top},
\end{align*}
where $\tilde{\bD}(\lambda)=\mbox{diag}\{s_k(d_k,\lambda)d_k,k=1,...,\bar{r}\}$. Once again using similar matrix algebra we arrive at a simpler expression for the degrees of freedom for the general class of reduced rank models

\begin{align}
\tilde{df}(\lambda) = tr\left\{\frac{\pal \mbox{vec}\{\bU\tilde{\bD}(\lambda)\bV^{\top}\}}{\pal \mbox{vec}(\bH)}\right\} = tr\left\{\frac{\pal \mbox{vec}\{\tilde{\bH}(\lambda)\}}{\pal \mbox{vec}(\bH)}\right\},\label{wdfdef}
\end{align}
where $\tilde{\bH}(\lambda)=\bU\tilde{\bD}(\lambda)\bV^{\top}$. It is now clear that the problem boils down to determining the divergence of a low-rank approximation of the matrix $\bH$ with respect to $\bH$ itself. This involves the derivatives of its singular values and singular vectors. Note that the singular values and vectors of a matrix are not only highly non-linear functions of the underlying matrix, they are also discontinuous on certain subsets of matrices \citep{ONeil2005}. This makes that degrees of freedom calculation for the reduced rank regression is a rather challenging problem. \citet{Stein1973} used derivatives of the singular values of a positive semi-definite matrix to estimate the risk improvement for a class of estimators for the mean of a multivariate Gaussian distribution. \citet{Tsukuma2008} also used a similar method to prove minimaxity of Bayes estimators for the mean matrix of a Gaussian distribution. We note that our set-up is very different from the ones considered by \citet{Stein1973} and \citet{Tsukuma2008}.  Specifically, we consider a regression setting where the design matrix makes the derivation more challenging.  Also as we aim to estimate the degrees of freedom of the model we need the derivatives of both singular values and vectors to compute the right hand side of \eqref{wdfdef}. There has also been a considerable amount of work on the smoothness and differentiability of the singular value decomposition of a real matrix in applied mathematics literature; main references include \citet{Magnus1998}, \citet{ONeil2005} and \citet{deLeeuw2007}. In view of this, we will proceed in two main steps: \\
\begin{enumerate}
\item Derive the partial derivatives in \eqref{df} and \eqref{wdfdef} under the condition that $\bH$ does not have repeated singular values, i.e., $d_1 > d_2 > \cdots >  d_{\bar{r}} > 0$. Use them to obtain an explicit exact unbiased estimator of degrees of freedom. 

\item Prove that the set where the partial derivatives do not exist has Lebesgue measure 0. \\
\end{enumerate}
\noindent The following two sections will address the aforementioned steps respectively and thus will complete the derivation of degrees of freedom estimator for the reduced rank estimators for multivariate regression under the SURE framework.

\section{Proposed estimator}\label{sec:ClosedForm}

We start by examining the derivatives of the singular values and singular vectors of a matrix with respect to an entry of the matrix itself. All the proofs are provided in the Appendix.

\begin{thm}\label{th:deriv}
Suppose $\bH$ is an $r_x\times q$ matrix of rank $q$, with $r_x\geq q$. Let its SVD be given by $\bH=\bU\bD\bV\t$, where $\bU\in \mathbb{R}^{r_x\times q}$, $\bU\t\bU=\bI$, $\bV\in \mathbb{R}^{q\times q}$, $\bV\t\bV=\bI$, and $\bD=\diag\{d_i,i=1,...,q\}$ with $d_1>\cdots >d_{q}>0$. Then for each $1\leq i\leq r_x$, $1\leq j\leq q$, and $1\leq k\leq q$,
\begin{align}
\frac{\partial \bv_{k}}{\partial h_{ij}}=&-(\bH\t\bH-d_k^2\bI)^{-}(\bH\t\bZ^{(ij)}+\bZ^{(ij)\top}\bH)\bv_k,\label{eq:deri1}\\
\frac{\partial d_{k}}{\partial h_{ij}}=&\frac{1}{2d_k}\bv_k\t(\bH\t\bZ^{(ij)}+\bZ^{(ij)\top}\bH)\bv_k,\label{eq:deri2}
\end{align}
where $(\bH\t\bH-d_k^2\bI)^{-}=\bV(\bD^2-d_k^2\bI)^+\bV^\top$ with $(\cdot)^+$ denoting the Moore-Penrose inverse, and $\bZ^{(ij)}=\partial \bH/\partial h_{ij}$ is an $r_x\times q$ matrix of zeros with only its $(i,j)$th entry being one.
\end{thm}

\noindent Without loss of generality, we have assumed $r_x \geq q$ in the above theorem. When $r_x \leq q$, the same results could be presented for $\bH\t$ with exchanged $r_x$ and $q$. Theorem \ref{th:deriv} is established from the general results in \citet{deLeeuw2007} about the derivatives of a generalized eigen-system. To ensure the derivatives are well-defined, we have assumed that the singular values are distinct. This is merely a restriction for real applications, as the observed singular values rarely coincide, a formal proof is provided in the next section. \\

\noindent It is not immediately clear whether the derived unbiased estimators in \eqref{df} and \eqref{wdfdef} may admit explicit form. Examining the SVD structure of $\bH$ sheds light on this problem. The pairs of singular vectors $(\bu_k,\bv_k)$ are orthogonal to each other, representing distinct directions in $\mathbb{R}^{r_x\times q}$ without any redundancy. Intuitively, these directions themselves are not distinguishable from each other, and their relative importance or contribution in constituting the matrix $\bH$ are entirely revealed by the singular values. This suggests that the complexity of reduced-rank estimation, as reflected by the relative complexity of a low rank approximation $\bH(r)$ or $\tilde{\bH}(\lambda)$ with respect to $\bH$, may only depend on the singular values of the matrix $\bH$ and the mechanism of singular-value shrinkage or thresholding. This is the main intuition that motivated the findings for explicit forms of \eqref{df} and \eqref{wdfdef}, which are summarized in the following theorems.

\begin{thm}\label{th:df1}
Let $\what{\bY}$ be the least squares estimator in \eqref{ols}. Let $\bar{r}=\mbox{rank}(\what{\bY}) = \min(r_x,q)$ and suppose the singular values of $\what{\bY}$ satisfy $d_1 > \cdots > d_{\bar{r}} > 0$. Consider the reduced-rank estimator $\what{\bY}(r)$ in \eqref{RRRsol}. An unbiased estimator of the effective degrees of freedom is
\begin{align*}
\what{df}(r)
=\left\{ \begin{array}{ll}
         \displaystyle \max(r_x,q)r + \sum_{k=1}^{r}\sum_{l=r+1}^{\bar{r}} \frac{d_k^2+d_l^2}{d_k^2-d_l^2},
         & r<\bar{r};\\
        r_x q,
         & r=\bar{r}.\end{array} \right.
\end{align*}
\end{thm}

\noindent The results are further generalized to the class of reduced-rank estimators in \eqref{ytilde}. It is worth noting that the weights $s_k(d_k,\lambda)$ are treated as random quantities since they are usually some functions of the singular values.

\begin{thm}\label{th:df2}
Let $\what{\bY}$ be the least squares estimator in \eqref{ols}. Let $\bar{r}=\mbox{rank}(\what{\bY})=\min(r_x,q)$ and suppose the singular values of $\what{\bY}$ satisfy $d_1 > \cdots > d_{\bar{r}} > 0$. Consider the reduced-rank estimator $\tilde{\bY}(\lambda)$ in \eqref{ytilde}, and let $\tilde{r}=\tilde{r}(\lambda)=\max\{k: s_k(d_k,\lambda)> 0.\}$. An unbiased estimator of the effective degrees of freedom is
\begin{equation*}
\tilde{df}(\lambda)
=\left\{ \begin{array}{ll}
        \displaystyle\max(r_x,q)\sum_{k=1}^{\tilde{r}}s_k + \sum_{k=1}^{\tilde{r}}\sum_{l=\tilde{r}+1}^{\bar{r}}\frac{s_k(d_k^2+d_l^2)}{d_k^2-d_l^2} + \sum_{k=1}^{\tilde{r}}\sum_{l\neq k}^{\tilde{r}}\frac{d_k^2(s_k-s_l)}{d_k^2-d_l^2}+\sum_{k=1}^{\tilde{r}}d_ks_k',
         & \tilde{r}<\bar{r};\\
        \displaystyle\max(r_x,q)\sum_{k=1}^{\tilde{r}}s_k
        + \sum_{k=1}^{\tilde{r}}\sum_{l\neq k}^{\tilde{r}}\frac{d_k^2(s_k-s_l)}{d_k^2-d_l^2}+\sum_{k=1}^{\tilde{r}}d_ks_k',
         & \tilde{r}=\bar{r}.
         \end{array} \right.
\end{equation*}
where for simplicity we write $s_k=s_k(d_k,\lambda)$ and $s_k'=\pal s_k(d_k,\lambda)/\pal d_k$.
\end{thm}

\noindent The explicit formulae presented in the above theorems facilitate further exploration of the behaviors and properties of the degrees of freedom.  For example, consider the unbiased estimator for reduced rank regression in Theorem \ref{th:df1}. It is always true that

\begin{align}
\what{df}(r)\geq \max(r_x,q)r + \sum_{k=1}^{r} \sum_{l=r+1}^{\bar{r}} \frac{d_k^2+0}{d_k^2-0}=(r_x+q-r)r, \qquad r=1,...,\bar{r}.\label{vsnaive}
\end{align}
This suggests that the proposed estimator is always greater than the naive estimator, i.e., the number of free parameters $(r_x+q-r)r$. Similar to the lasso method in univariate regression problems \citep{Tibshirani1996, Zou2007}, the reduced-rank estimation can be viewed as a latent factor selection procedure, in which we both construct and search over as many as $\bar{r}$ latent linear factors. Therefore, the increments in the degrees of freedom as shown in \eqref{vsnaive} can be interpreted as the price we have to pay for performing this latent factor selection. For the general methods considered in Theorem \ref{th:df2}, this inequality no longer holds, due to the shrinkage effects induced by the weights $0\leq s_k \leq 1$. The reduction in the degrees of freedom due to singular-value shrinkage can offset the price paid for searching over the set of latent variables. Therefore, similar to lasso, adaptive singular-value penalization can provide effective control over the model complexity \citep{TibRyan2012, Chen2012}.\\

\noindent Although the unbiased estimator and the naive estimator are quite different, some interesting connections can be made. The two estimators are close to each other when they are evaluated at the true underlying rank, especially when the signal is strong relative to the noise level. This phenomenon was also noted in the empirical studies. Suppose the true model rank is $\mbox{rank}(\bB) = r^*$. Intuitively, the $\bar{r}-r^*$ smallest singular values from least squares may be close to zero and are not comparable to the $r^*$ largest ones; using the approximation $d_k\approx 0$, $k=r^*+1,...,\bar{r}$, we obtain $\what{df}(r^*)\approx (r_x+q-r^*)r^*$. A more rigorous argument can be made from either classical or high-dimensional theoretical perspective. In classical large $n$ settings, under standard assumptions, the consistency of the least squares estimation can be readily established \citep{Reinsel1998}. Using techniques such as the perturbation expansion of matrices \citep{Izenman1975}, the consistency of $\what{\bY}$ implies the consistency of the estimated singular values, i.e., the first $r^*$ estimated singular values converge to their nonzero true counterparts while the rest converge to zero in probability. It follows that
\begin{equation}
\what{df}(r^*)\rightarrow_p (r_x + q - r^*)r^*
\end{equation}
\noindent in probability as $n\rightarrow \infty$. An immediate implication of this result is that for each $r=1,...,\bar{r}$, if we assume the true model is of rank $r$, then in an asymptotic sense, the number of free parameter, $(r_x+q-r)r$, is the correct degrees of freedom to use. This clearly relates to the error degrees of freedom of the classical asymptotic $\chi^2$ statistic from the likelihood ratio test of $H_0: \mbox{rank}(\bB) = r$ \citep{Izenman1975}, for each $r=1,...,\bar{r}$. In high-dimensional models, non-asymptotic prediction error bounds have been developed for the considered reduced-rank estimation methods, and the minimax convergence rate in fact coincides with the number of free parameters \citep{Rohde2011, Bunea2011, Chen2012}. These results provide further justification of the proposed unbiased estimator and reveal the limitations, the underlying assumptions and the asymptotic nature of the naive estimator. \\

\noindent The derived formulae also reveal some interesting behaviors of rank reduction. In essence, the reduced-rank methods distinguish the signal from the noise by examining the estimated singular values from least squares estimation: the large singular values more likely represent the signals while the small singular values mostly correspond to the noise \citep{Bunea2011, Chen2012}. By rank reduction, we aim to recover the signals exceeding certain noise level. Consider the case when $d_k$ and $d_{k+1}$ are close for some $k=1,...,\bar{r}-1$. It can be argued that the true model rank is unlikely to be $k$, because the $(k+1)$th layer and the $k$th layer are hardly distinguishable. Indeed, this is reflected from the degrees of freedom: for $r=k$, the formula includes a term $(d_k+d_{k+1})/(d_k-d_{k+1})$, which can be excessively large. On the other hand, there is no such term for $r=k+1$. Consequently, the unbiased estimator of the degrees of freedom may not monotonically increase as the rank $r$ increases, in contrast to the naive estimator. In the above scenario, the estimates for $r=k$ can even be larger than that of $r=k+1$. This automatically reduces the chance of $k$ being selected as the final rank.

\section{Existence of partial derivatives almost everywhere}\label{sec:Existence}

\noindent One of the main technical assumptions for Stein's degrees of freedom estimator is that the partial derivatives must exist almost everywhere. Theorem \ref{th:deriv} gives us the condition, $d_1 > d_2 > \ldots > d_{\bar{r}} > 0$ for the existence of the partial derivatives of singular values and singular vectors of $\mbf{H} \in \mathbb{R}^{r_x \times q}$, where $\{d_i\}_{i = 1}^{\bar{r}}$ denote the singular values of $\bH$. Also recall that $\bar{r} = \min(r_x, q)$. Therefore, to apply Stein's framework we must show that matrices with full rank and non-repeated singular values are \emph{``dense"} in the set of all real matrices of dimension $r_x \times q$. The following theorem gives that result.
\begin{thm}\label{thm1}
Let $\mathbb{R}^{r_x \times q}$ be the space of all real-valued $r_x \times q$ dimensional matrices equipped with the Lebesgue measure $\mu$. Also, let $\mathcal{S} \subseteq \mathbb{R}^{r_x \times q}$ denote the subset of matrices that have full rank and no repeated singular values. Then $\mu( \mathcal{S}) = 1$.
\end{thm}

\noindent To prove the theorem, we start with a few definitions and facts from algebraic geometry and matrix analysis.

\begin{defn}\label{defn1}
An algebraic variety over $\mathbb{R}^k$(or $\mathbb{C}^k$) is defined as the set of points satisfying a system of polynomial equations $\left\{f_{\ell}(x_1, x_2, \ldots, x_k) = 0; \ \ \ell \in \mathcal{I} \right\}$.
\end{defn}

\noindent Here each $f_{\ell}(\cdot)$ is a polynomial function of its arguments and $\mathcal{I}$ denotes an index set. If at least one of the $f_{\ell}(\cdot) \not\equiv 0$, then it is called a proper sub-variety. Note that a proper sub-variety must be of dimension less than $k$ and therefore has Lebesgue measure 0 in $\mathbb{R}^k$ \citep{Allman2009}.  For a more detailed discussion, we recommend \citet{Hartshorne1977} or \citet{Cox2007}. 


\begin{prop}\label{P2} \citep{Laub2004}
Any square symmetric matrix $\mbf{M} \in \mathbb{R}^{k \times k}$ has at least one repeated eigenvalue if and only if $rank\left(\mbf{M} \otimes \mbf{I}_k - \mbf{I}_k \otimes \mbf{M}\right) < (k^2 - k)$.
\end{prop}

\noindent Now we prove the theorem.
%
First we define
\begin{eqnarray*}
\mathcal{S}_1 &=& \{\mbf{A} \in \mathbb{R}^{r_x\times q} : \mbf{A} \textrm{ has at least one 0 singular value}\}, \\
\mathcal{S}_2 &=& \{\mbf{A} \in \mathbb{R}^{r_x\times q} : \mbf{A} \textrm{ has at least one repeated singular value}\}.
\end{eqnarray*}

\noindent Note that $\mathcal{S}^c = \mathcal{S}_1 \cup \mathcal{S}_2$, thus it is enough to show that $\mu(S_1) = 0$ and $\mu(S_2) = 0$. By definition \ref{defn1} and the discussion above it suffices to show that $\mathcal{S}_1$ and $\mathcal{S}_2$ are proper sub-varieties of $\mathbb{R}^{r_x\times q}$. Note that $\mathcal{S}_1$ can be rewritten as follows
\[
\mathcal{S}_1 = \{\mbf{A} \in \mathbb{R}_{r_x \times q}: det (\mbf{A}\t\mbf{A}) = 0\}.
\]

\noindent Here $det(\cdot)$ denotes the determinant operator for a square matrix.  Note that $det(\mbf{A}\t\mbf{A})$ is a non-trivial polynomial in entries of $\mbf{A}$ and hence $\mathcal{S}_1$ is a proper sub-variety and has Lebesgue measure 0.  For $\mathcal{S}_2$ note that if $\mbf{A} \in \mathbb{R}^{p \times q}$ has at least one repeated singular value, it implies that $\mbf{A}^T\mbf{A} \in \mathbb{R}^{p \times q}$ has at least one repeated eigenvalue.  Then in view of proposition \ref{P2}, $\mathcal{S}_2$ can be reformulated as
\[
\mathcal{S}_2 = \left\{\mbf{A} \in \mathbb{R}_{r_x \times q}: rank\left(\mbf{A}\t \mbf{A} \otimes \mbf{I}_q - \mbf{I}_q \otimes \mbf{A}\t\mbf{A}\right) < (q^2 - q)\right\}.
\]

\noindent This is an algebraic variety since it can be expressed as the solution to all minors of order $\geq (q^2 - q)$ being equal to 0, which are all polynomial equations in the entries of $\mbf{A}$. Thus, we have shown that, $\mu(\mathcal{S}_1 \cup \mathcal{S}_2) = 0$.

\section{Simulation studies}\label{sec:Sim}

In this section, we evaluate the performance of the proposed method by simulation studies.  Specifically, we aim to demonstrate two things: 1) the exact unbiased estimator of the degrees of freedom for the reduced rank regression is in general significantly higher than the naive estimator; 2) using the exact estimator of the degrees of freedom enables us to gain prediction accuracy over the naive estimator.

\subsection{Unbiasedness}

In this simulation, we aim to show that the degrees of freedom estimator defined via Theorem \ref{th:df1} is unbiased and it can be significantly higher than the naive estimator that simply counts the number of free parameters. Here unbiasedness is defined over the error distribution, and we treat $\mbf{X}$ as a fixed design matrix. We conduct the study at two different parameter settings one for low-dimension and one for high-dimension. Parameters of the setting are as follows
\begin{eqnarray*}
\mbox{Setting I} & : & n= 100, p = 20, q = 12, r_0 = 6 \\
\mbox{Setting II} & : & n = 40, p = 80, q = 50, r_0 = 10
\end{eqnarray*}
where $r_0$ denotes the true rank of $\mbf{B}$. Let $\mbf{\Sigma}$ denote the covariance matrix of the predictor variables, $\mbf{X}$, and we set
$\mbf{\Sigma}_{jj'} = 0.3^{\lvert j-j' \rvert}$.  Rows of the predictor matrix are generated independently from $N_p(\mbf{0}, \mbf{\Sigma})$.  To control the singular structure of $\mbf{B}$ through the covariance of signals $\mbf{XB}$, $\mbf{B}^T\mbf{\Sigma B}$, we take the left singular vectors of $\mbf{B}$ the same as the eigenvectors of $\mbf{\Sigma}$, whereas the right singular vectors of $\mbf{B}$ are generated by orthogonalizing a random standard normal matrix. The difference between successive non-zero singular value of $\mbf{B}$ is fixed at $2$.  The error matrix is generated from i.i.d. standard normal distribution.  We replicate the process $200$ times; note that the design matrix remains fixed.  We compare the proposed exact method against the data perturbation technique \citep{Ye1998} and the Monte-Carlo estimator of the true degrees of freedom which is computed from \eqref{dfOrig}.  For the data perturbation method, we consider $50$ perturbations of the response matrix for each replication to estimate the partial derivatives numerically.  We used the choice of $0.1 \sigma$ for the perturbation size, where $\sigma$ is the error standard deviation. Ideally we would expect the proposed exact estimator to be fairly close to the data perturbation and Monte-Carlo estimator on average.  We compare estimators against the naive degrees of freedom estimate namely, $df_n(r) = r(r_x + q - r)$, which denotes the number of free parameters in a $p \times q$ matrix of rank $r$.  Note that the naive estimator does not depend on the data. \\

\begin{figure}[h!]
\begin{center}
\includegraphics[width=6in]{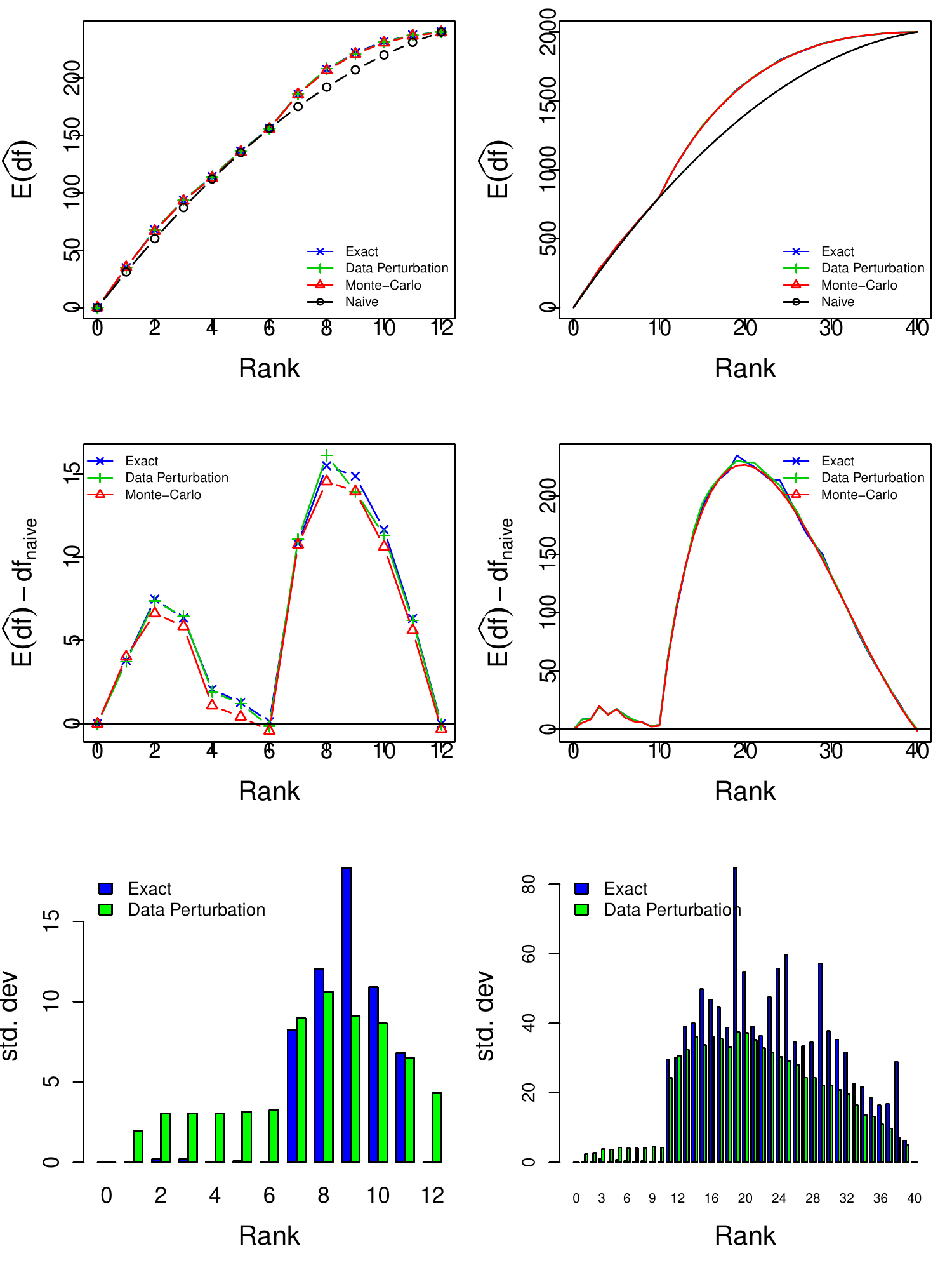}
\caption{Left column: low dimensional setting, right column: high-dimensional setting.  Top row: average of estimated degrees of freedom over 200 replications.  Middle row: difference between the estimated degrees of freedom and the naive degrees of freedom. Bottom row: standard error of the estimated degrees of freedom.}
\label{fig:DF}
\end{center}
\end{figure}

\noindent On the top row of Figure \ref{fig:DF} we see that for both high-dimensional and low-dimensional settings the proposed exact method, the data perturbation estimator and the Monte-Carlo estimator are nearly identical; further, they are significantly higher than the naive estimator, as indicated in the middle row of Figure \ref{fig:DF}. The difference is especially large once we go above the correct rank. It also justifies our theoretical intuition that the exact estimators seem to match the naive estimator very closely at the true rank. The bottom panels allow us to get a sense of the variability of the estimation procedures. Standard error for the exact method is orders of magnitudes smaller than that of data perturbation below the true rank but once we go above the true rank the standard errors of the exact estimator becomes drastically higher. This arises from the fact that once we go above the true rank, the singular values of $\widehat{\mbf{Y}}$ basically correspond to noise, and can be very close to each other.  Hence slight perturbations of the data might lead to different singular directions being selected, which implies higher variability in model complexity.  This has also been noted by \citet{Ye1998}, that is, if we are trying to fit pure error components, the degrees of freedom tends to be higher and unstable. \\

\subsection{Prediction performance}

The previous set of simulations have shown that the exact degrees of freedom estimator can be significantly different from the number of free parameters estimator.  Degrees of freedom estimates are commonly used in various model selection criteria.  In this subsection, we aim to show that for reduced rank regression, we can gain in prediction accuracy by using the exact degrees of freedom estimator in a model selection criterion instead of the naive estimator. Since our focus is on prediction accuracy, we consider generalized cross-validation(GCV) \citep{Golub1979} as our model selection criterion. This choice was motivated by the fact that it does not require an estimate for the error variance. Other popular choices such as Mallows $C_p$ \citep{Mallow1973} require an estimate of error variance which is hard to obtain in high-dimensional settings.  In the context of reduced rank regression, the GCV criterion is defined as follows
\begin{eqnarray*}\label{MSC}
GCV(r) &=& \frac{nq \lVert \mbf{Y} - \widehat{\mbf{Y}}(r)\rVert_F^2}{(nq - df(r))^2}.
\end{eqnarray*} 

\noindent We select the model that minimizes the GCV criterion over $1\leq r \leq \min(n, p, q)$. Once again we choose a low-dimensional and a high-dimensional setting for a comprehensive comparison. 
\begin{eqnarray*}
\mbox{LD Setting} & : & n= 50, p = 12, q = 10, r_0 = 3 \\
\mbox{HD Setting} & : & n = 40, p = 80, q = 50, r_0 = 5
\end{eqnarray*}

\noindent For each setting we consider two different levels for error variance, namely, $\sigma^2=1$ and $4$. This allows us to controls the signal to noise ratio defined as $\mbox{SNR} = d_{r_0}(\mbf{XB})/ d_1(\mbf{E})$.  The numerator stands for the smallest non-zero singular value of the signal matrix, a measure of the signal strength, whereas the largest singular value of the error matrix measures the noise strength \citep{Bunea2011}. Correlation among predictor variables is kept at a moderate level of $0.5$. The data generation scheme remains the same as before. We fit the optimal model based on GCV with the exact degrees of freedom (GCV(e)) and GCV with the naive degrees of freedom (GCV(n)) and report the following: estimation error $\mbox{Est} = 100 \lVert \mbf{B} - \widehat{\mbf{B}}\rVert_F^2/(pq)$, the prediction error $\mbox{Pred} = 100 \lVert \mbf{XB} - \mbf{X}\widehat{\mbf{B}}\rVert_F^2/(nq)$ as well as the selected rank. .   Table \ref{tab:pp} summarizes the results. We report the averages over $100$ replications and the numbers inside the parenthesis indicate standard error. \\

\begin{table}[h]
\caption{Prediction performance  comparison between different model selection criteria}
\label{tab:pp}
\begin{center}
\begin{tabular}{c| c| c c| c c}
\hline
Error Variance & Performance & 
\multicolumn{2}{c|}{LD setting}&\multicolumn{2}{c}{HD setting} \\
and SNR& Measure& GCV(e) & GCV(n) & GCV(e) & GCV(n) \\
\hline
\multirow{3}{*}{$\sigma^2 = 1$, $\mbox{SNR} \approx 1$} & Est & 1.56(0.4) & 1.80(0.8) & 3.25(0.5) & 3.30(0.5) \\
& Pred & 11.95(2.2) & 12.97(3.4) & 22.89(1.5) & 28.28(4.3) \\
& Rank & 3.01(0.1) & 3.18(0.4) & 4.84(0.4) & 5.30(0.5) \\
\hline
\multirow{3}{*}{$\sigma^2 = 4$, $\mbox{SNR} \approx 0.5$} & Est & 6.00(2.7) & 7.47(3.4) & 3.77(0.5) & 4.00(0.6) \\
& Pred & 50.64(10.8) & 54.31(10.8) & 78.48(6.2) & 89.93(17.4) \\
& Rank & 2.41(0.6) & 2.86(0.6) & 4.00(0.0) & 4.46(0.6) \\
\hline
\end{tabular}
\end{center}
\end{table}

\begin{figure}[h!]
\begin{center}
\includegraphics[width=3in, height=2.5in]{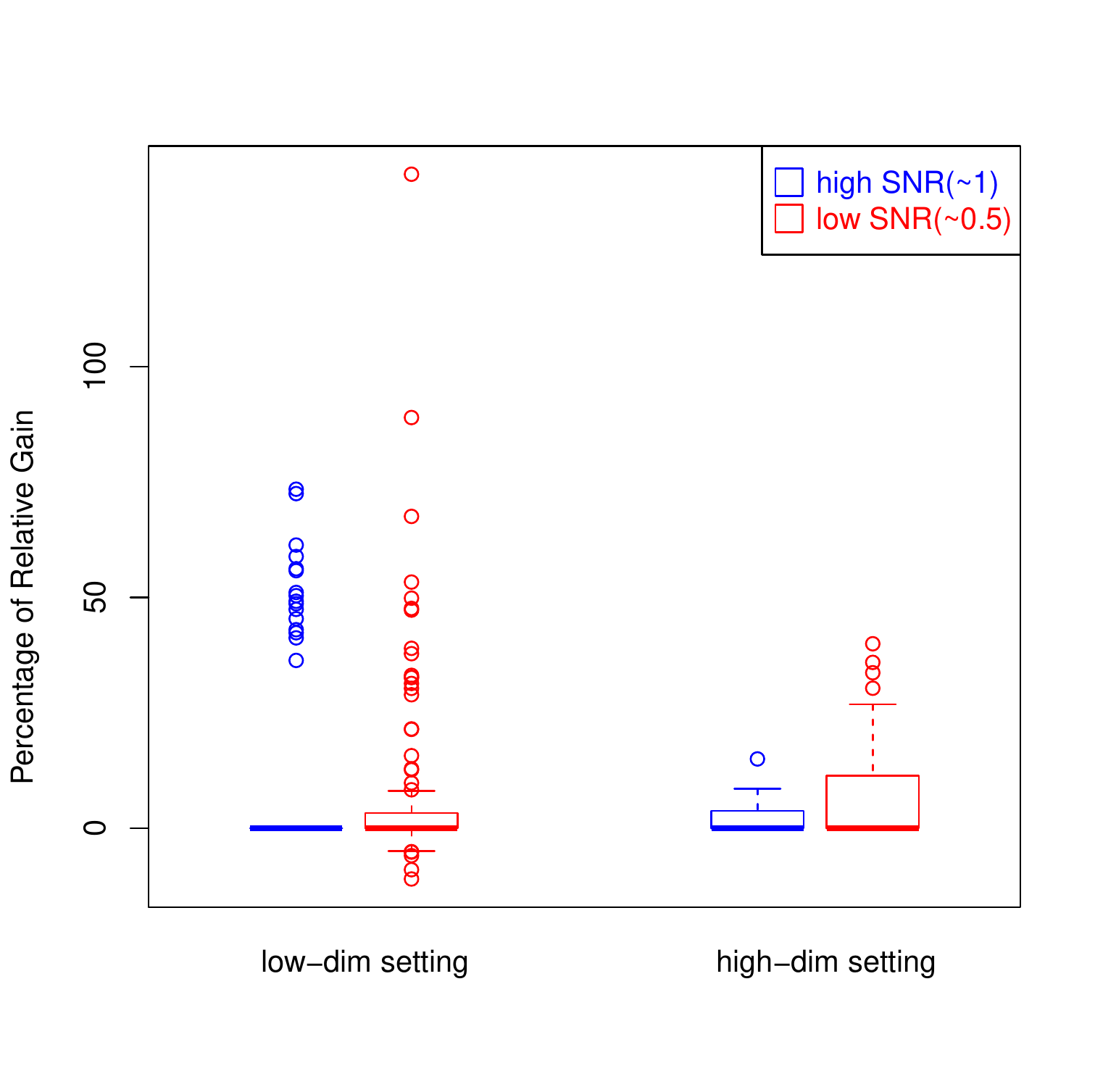}
\caption{Relative gain in prediction error by using the exact estimator of degrees of freedom over the naive estimator}
\label{fig:PE}
\end{center}
\end{figure}

\noindent We find that using the proposed exact degrees of freedom estimator in GCV criterion performs better in terms of prediction accuracy than its naive counterpart.  It has lower average estimation error and prediction error for all the settings. The relative gain is larger for the prediction error. We wish to note that similar results were obtained at other levels of correlation but were excluded to facilitate brevity. For the low-dimensional setting where an estimator of $\sigma^2$ is available we also studied the performance of Mallow's $C_p$ criterion and once again the results were very close to the ones reported and therefore excluded. We find that in the settings with moderately high SNR, the naive degrees of freedom estimator tends to overestimate the rank leading to inflated error measures. On the other hand in low SNR settings often the smallest non-zero singular values have very little explanatory power and therefore selecting a lower rank model enables us to do better in terms of prediction accuracy due to the bias-variance trade-off. As the exact degrees of freedom estimator is usually higher than the naive estimator it penalizes more strictly and selects a simpler model which predicts better. To get a better understanding for the comparison between the two degrees of freedom estimators, we also computed the percentage of pairwise relative gain, which is defined as follows

\begin{equation*} 
\mbox{PRG} = 100 \times \frac{(\mbox{Pred(n)} - \mbox{Pred(e)})}{\mbox{Pred(e)}}\%, 
\end{equation*}
\noindent where Pred(e) denotes the prediction error when using exact degrees of freedom estimator in the GCV criterion, similarly Pred(n) denotes the prediction error when using the naive degrees of freedom estimator in GCV. Note that these ratios are computed on a per data set basis.  As we can see in Figure \ref{fig:PE}, the boxplots tend to stay above zero almost always indicating that the exact degrees of freedom outperforms the naive estimator consistently. Also the relative gain is larger in the high-dimensional scenario.

\section{Analysis of Arabidopsis thaliana data}\label{sec:Dat}

In this section, we apply the proposed degrees of freedom methodology to fit a reduced rank model to a genetic association data set that was published in \citet{Wille2004}.  This is a microarray experiment aimed at understanding the regulatory control mechanisms between the isoprenoid gene network in Arabidopsis thaliana plant (more commonly known as thale cress or mouse-ear cress).  It is known that isoprenoids serve many important biochemical functions in plants.  To monitor the gene-expression levels, 118 GeneChip microarray experiments were carried out.  The predictors consist of 39 genes from two isoprenoid bio-synthesis pathways namely MVA and MEP, whereas the responses consist of gene-expression of 795 genes from 56 metabolic pathways, many of which are downstream of the two pathways considered as predictors.  Thus some of the responses are expected to show significant associations to the predictor genes.  To facilitate it further, we select two downstream pathways namely, Caroteniod and Phytosterol as our responses.  It has already been proven experimentally that the Carotenoid pathway is strongly attached to the MEP pathway, whereas the Phytosterol pathway is significantly related to the MVA pathway.  See \citet{Wille2004} and the references therein for a more detailed discussion on the biological aspects.  Finally we have 118 observations on $p=39$ predictors and $q=36$ responses.  All the predictors and responses are log-transformed to reduce the skewness of the data.  We also standardize the responses in order to make them comparable.  \\

\noindent 
We split the data set randomly into training and test sets of equal size. The model is fit using the training samples and then we use it to predict on the test set. The performance measure under consideration is the usual mean squared prediction error
\begin{equation}\label{PerfM}
\displaystyle \mbox{MSPE} = \frac{2}{nq} \lVert \mbf{Y}_{\mbox{test}} -  \widehat{\mbf{Y}}_{\mbox{test}}\rVert^2_F.
\end{equation}

\noindent The entire process is repeated 100 times based on random splits to ensure that the results remain robust to the process of splitting.  We used Mallow's $C_p$, GCV and BIC with the exact degrees of freedom and the naive degrees of freedom to select the optimal rank. \\

\begin{figure}[h!]
\begin{center}
\includegraphics[width = 6in, height = 2.5in]{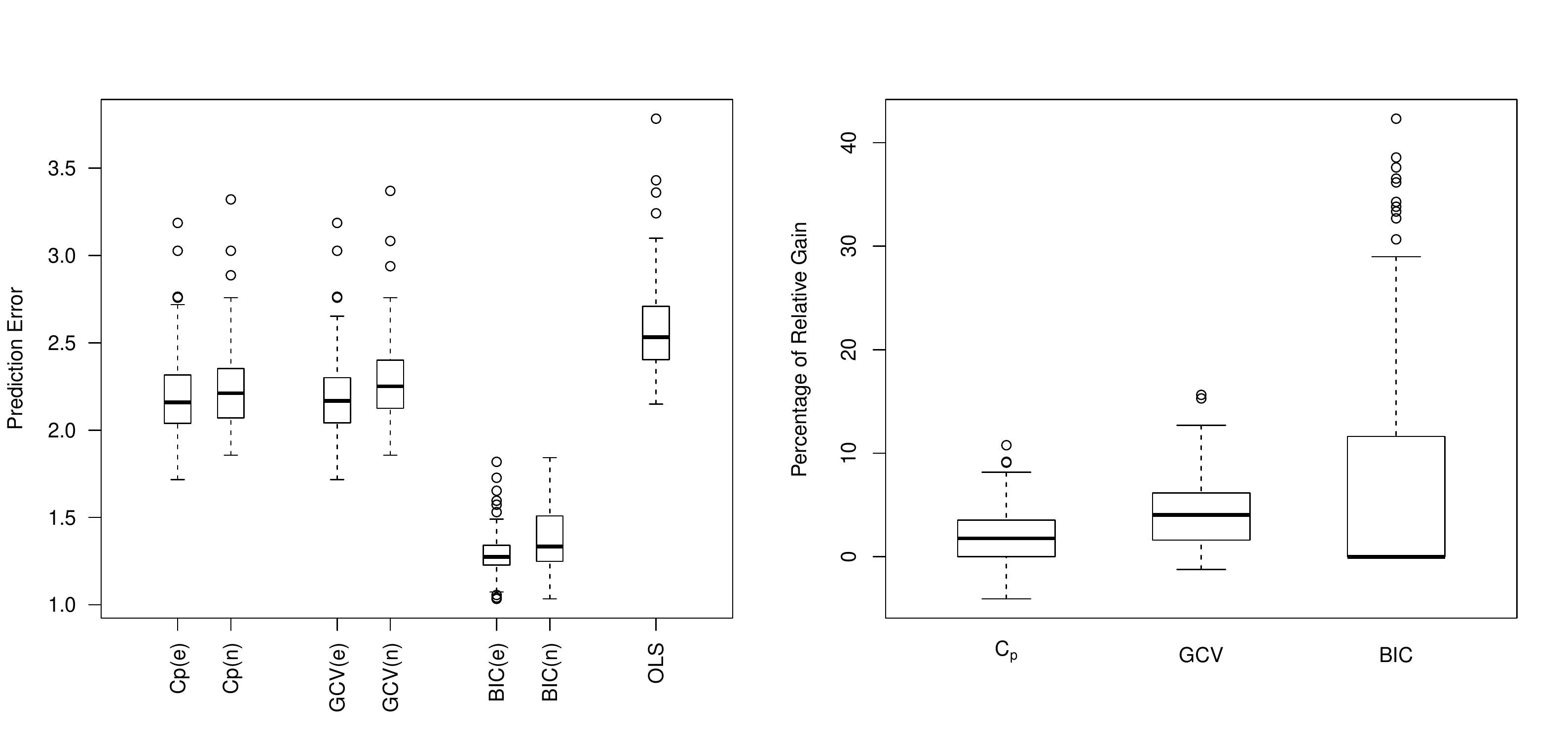}
\caption{Left: boxplot of mean square prediction error of each method over 100 random splits; Right: Relative increase in prediction error for using naive degrees of freedom over the exact degrees of freedom estimator for each model selection criteria.}
\label{fig:box}
\end{center}
\end{figure}

\begin{table}[h]
\caption{Prediction accuracy and rank selection performance for the competing methods on the Arabidopsis thaliana data.}
\label{tab:da}
\centering
\begin{tabular}{c | c c c c c c c}
\hline
& Cp(e) & Cp(n) & GCV(e) & GCV(n) & BIC(e) & BIC(n) & OLS \\
\hline
Avg(Pred Err) & 2.197 & 2.243 & 2.192 & 2.282 & 1.297 & 1.387 & 2.589 \\ 
Std(Pred Err) & 0.250 & 0.246 & 0.248 & 0.246 & 0.134 & 0.201 & 0.282 \\ 
Mean(Est Rank) & 8.760 & 9.710 & 8.680 & 10.520 & 1.090 & 1.480 & -- \\
Std(Est Rank) & 1.15 &  0.83 &  1.27 &  0.97 &  0.38 & 0.76 & --\\
\hline
\end{tabular}
\end{table}

\noindent The mean squared prediction errors for each method are summarized using the boxplot in Figure \ref{fig:box}.  As we can see, for all three model selection criteria considered, the use of the exact unbiased estimator enables us to outperform the one which uses the naive estimator in terms of prediction accuracy. The relative gain is almost always positive as we can see from the right panel of Figure \ref{fig:box}. Also among the three model selection criteria BIC appears to be the clear winner in terms of prediction error by selecting a very parsimonious model (Table \ref{tab:da}). 

\section{Concluding remarks}\label{sec:Sum}

We have proposed an exact unbiased estimator of the degrees of freedom for a general class of reduced rank estimators for the multivariate linear regression model in the framework of SURE. The proposed estimator can be computed explicitly leading to an efficient model selection procedure compared to computationally expensive cross-validation or data-perturbation based methods. The closed form also provides us with some much needed insight regarding the connection between the exact and the naive degrees of freedom estimator. The proposed methodology does not make any assumption regarding the dimensions of the problem or the rank of the design matrix and is very suitable for application to high-dimensional problems $(p, q > n)$ as illustrated via several numerical examples. The methods developed here are quite general and can be extended to other related estimation procedures that employ regularization of the singular values, e.g., reduced rank ridge regression \citep{Mukherjee2011}. There are several directions for future research. We have mainly considered the reduced-rank estimators which share the same set of singular vectors with the least squares solution. It would be interesting and challenging to extend the results for other reduced-rank methods, such as, the nuclear-norm penalized regression \citep{Yuan2007}. Since reduced-rank estimation can be more effective when combined with sparse estimation, e.g., selecting latent factors of a sparse subset of original variables, it would be very interesting to extend the methodology to sparse and low-rank models \citep{Zou2007, Chen2012JRSSB, Bunea2012}. Another pressing problem concerns investigating the proposed approach in reduced rank generalized linear models \citep{Yee2003, Li2007b, She2012}. Finally, as the reduced rank methods are commonly used in multiple time series analysis, the proposed approach can be extended to these settings, including reduced rank models with multiple sets of regressors \citep{Velu1991} and the co-integration problem \citep{Anderson2002a}. 

\nolinenumbers
\bibliographystyle{asa}
\bibliography{refAshin}

%

\appendix
{\center\section*{Appendix}}

\subsection*{Derivation of Equation \eqref{df}}
Note that
\begin{eqnarray*}
\widehat{\mbf{Y}} &=& \mbf{XQS}^{-1}\bH, \\
\implies \widehat{\mbf{Y}}(r) & = & \mbf{XQS}^{-1}\bH(r), \quad r = 1, \ldots, \bar{r},
\end{eqnarray*}

\noindent Using the trace identity, $tr(\bA\bB) = tr(\bB\bA)$, the equality, $vec(\mbf{ABC}) = (\mbf{C}\t \otimes \mbf{A}) vec(\mbf{B})$ and the chain rule of differentiation we get

\begin{eqnarray} 
\widehat{df}(r) &=& tr \left\{ \frac{\partial vec(\widehat{\mbf{Y}}(r))}{\partial vec(\mbf{Y})}\right\} \nonumber \\
& = & tr \left\{ \left[\mbf{I}_q \otimes \bX\bQ\bS^{-1}\right]  \left(\frac{\partial vec(\bH(r))}{\partial vec(\mbf{Y})}\right)\right\} \nonumber \\
& = & tr \left\{ \left[\mbf{I}_q \otimes \bX\bQ\bS^{-1}\right]  \left(\frac{\partial vec(\bH(r))}{\partial vec(\bH)}\right)\left(\frac{\partial vec(\bH)}{\partial vec(\bY)}\right)\right\} \nonumber \\
& = & tr \left\{ \left[\mbf{I}_q \otimes \bX\bQ\bS^{-1}\right]  \left(\frac{\partial vec(\bH(r))}{\partial vec(\mbf{Y})}\right) \left[\mbf{I}_q \otimes \bS^{-1}\bQ\t\bX\right]\right\} \nonumber \\
&=& tr \left\{ \frac{\partial vec(\bH(r))}{\partial vec(\mbf{Y})}\right\}. \nonumber \nonumber
\end{eqnarray}

\subsection*{Proof of Theorem \ref{th:deriv}}

We acknowledge that the proof of Theorem \ref{th:deriv} is mainly based on the results developed in \citet{deLeeuw2007} about the derivatives of a generalized eigensystem. Note that we have assumed $r_x\geq q$, and the same results can be presented for $\bH\t$ when $r_x\leq q$. \\

\noindent Denote $\bA=\bH^\top\bH$, and let $(d^2,\bv)$ denote a pair of eigenvalue and eigenvector of $\bA$. Suppose $\bA$ is two times continuously differentiable at $\theta$, e.g., $\theta=h_{ij}$ for any $i=1,...,r_x$ and $j=1,...,q$. Then the eigenvalues and eigenvectors are also differentiable at $\theta$. From
$$
\bA\bv=d^2\bv,
$$
it follows that
$$
\frac{\pal \bA}{\pal \theta}\bv + \bA\frac{\pal \bv}{\pal \theta}=d^2\frac{\pal \bv}{\pal \theta}+\frac{\pal d^2}{\pal \theta}\bv,
$$
and this gives
\begin{align}
(\bA-d^2\bI)\frac{\pal \bv}{\pal \theta}=-(\frac{\pal \bA}{\pal \theta}-\frac{\pal d^2}{\pal \theta}\bI)\bv.\label{eq:eigen1}
\end{align}
Premultiplying both sides by $\bv^\top$ gives
$$
\bv^\top(\bA-d^2\bI)\frac{\pal \bv}{\pal \theta}=-\bv^\top\frac{\pal \bA}{\pal \theta}\bv +\frac{\pal d^2}{\pal \theta}.
$$
It is obvious that the left-hand-side equals to 0, and it then follows that
\begin{align}
\frac{\pal d}{\pal \theta}=\frac{1}{2d}\bv^\top\frac{\pal \bA}{\pal \theta}\bv.\label{eq:eigenvalue}
\end{align}

\noindent Define $(\bA-d^2\bI)^{-}=\bV(\bD^2-d^2\bI)^+\bV^\top$ with $(\cdot)^+$ denoting the Moore-Penrose inverse. Therefore, $(\bA-d^2\bI)^{-}(\bA-d^2\bI)=\bI-\bv\bv^\top$ and $(\bA-d^2\bI)^{-}\bv=0$. Premultiplying both sides of (\ref{eq:eigen1}) by $(\bA-d^2\bI)^{-}$ gives
$$
(\bI-\bv\bv^\top)\frac{\pal \bv}{\pal \theta}=-(\bA-d^2\bI)^{-}\frac{\pal \bA}{\pal \theta}\bv.
$$
From $\bv^\top\bv=1$, we know that $\bv^\top(\pal \bv/\pal \theta)=0$. It then follows that
\begin{align}
\frac{\pal \bv}{\pal \theta}=-(\bA-d^2\bI)^{-}\frac{\pal \bA}{\pal \theta}\bv.\label{eq:eigenvector}
\end{align}

\noindent Define $\bZ^{(ij)}=\partial \bH/\partial h_{ij}$ be an $r_x\times q$ matrix of zeros with only its $(i,j)$th entry equaling to one. For any $\theta=h_{ij}$,
\begin{align}
\frac{\pal \bA}{\pal h_{ij}}=\bH^\top\bZ^{(ij)}+\bZ^{(ij)\top}\bH.\label{eq:hderiv}
\end{align}
The proof is completed by combining the results in (\ref{eq:eigenvalue}), (\ref{eq:eigenvector}) and (\ref{eq:hderiv}).

\subsection*{Proof of Theorem \ref{th:df1}}

For simplicity and without loss of generality, we assume $r_x\geq q$. When $r_x\leq q$, one can repeat the same proof using $\bH\t$. When $r=q$, the result $\what{df}(q)=r_x q$ holds trivially. So in the following, we consider $r<q$. Consider $\partial\bH^{(r)}/\partial h_{ij}$ for any $1\leq i\leq r_x$, $1\leq j\leq q$. Because $\bH^{(r)}=\bH\sum_{k=1}^{r}\bv_k\bv_k\t$, by the chain rule, we have
\begin{align}
\frac{\pal\bH^{(r)}}{\pal h_{ij}}
=&\frac{\pal\bH}{\pal h_{ij}}\sum_{k=1}^{r}\bv_k\bv_k\t
+ \bH\sum_{k=1}^{r}\frac{\pal\bv_k}{\pal h_{ij}}\bv_k\t
+\bH\sum_{k=1}^{r}\bv_k\frac{\pal\bv_k\t}{\pal h_{ij}} \notag\\
=&\bZ^{(ij)}\bV^{(r)}\bV^{(r)\top} - \bH\sum_{k=1}^{r}\left\{(\bH\t\bH-d_k^2\bI)^{-}(\bH\t\bZ^{(ij)}+\bZ^{(ij)\top}\bH)\bv_k\bv_k\t\right\}\notag\\
&-\bH\sum_{k=1}^{r}\left\{\bv_k\bv_k\t(\bH\t\bZ^{(ij)}+\bZ^{(ij)\top}\bH)(\bH\t\bH-d_k^2\bI)^{-}\right\}.\label{eq:parij}
\end{align}

\noindent Consider the first term on the right-hand-side of (\ref{eq:parij}). Its $(i,j)$th entry equals to $\sum_{k=1}^{r}v_{jk}^2$. Therefore, its contribution to the degrees of freedom \eqref{df} is
\begin{align}
\sum_{i=1}^{r_x}\sum_{j=1}^{q}\sum_{k=1}^{r}v_{jk}^2=r_xr,\label{eq:term1}
\end{align}
because $\sum_{j=1}^{q}v_{jk}^2=1$. We know
$$
(\bH\t\bH-d_k^2\bI)^{-}=\sum_{l\neq k}^{q} \frac{1}{d_l^2-d_k^2}\bv_l\bv_l\t.
$$
We also have
\[\bH\t\bZ^{(ij)}+\bZ^{(ij)\top}\bH=
 \begin{pmatrix}
   &  & h_{i1} & &\\
   &  &  \vdots & &\\
  h_{i1} & \cdots & 2h_{ij} & \cdots & h_{iq}\\
   &  &  \vdots & &\\
   &  &  h_{iq} & &\\
 \end{pmatrix}.
\]

\noindent Now consider the second term on the right-hand-side of (\ref{eq:parij}). After some algebra, its $(i,j)$th entry can be written as $\bu_i\t\bD\ba^{(ij)}$, where $\ba^{(ij)}\in \mathbb{R}^{q}$ and
$$
a_k^{(ij)}=-\sum_{l\neq k}^{r}\frac{1}{d_k^2-d_l^2}(v_{jk}v_{jl}\bh_i\t\bv_l+v_{jl}^2\bh_i\t\bv_k),\qquad k=1,...,q.
$$

\noindent Similarly, the $(i,j)$th entry of the third term on the right-hand-side of (\ref{eq:parij}) is given by $\bu_i\t\bD\bb^{(ij)}$, where $\bb^{(ij)}\in \mathbb{R}^{q}$,
$$
b_k^{(ij)}=-\sum_{l\neq k}^{q}\frac{1}{d_l^2-d_k^2}(v_{jk}v_{jl}\bh_i\t\bv_s+v_{jl}^2\bh_i\t\bv_k),\qquad k=1,...,r,
$$
and $b_k^{(ij)}=0$ for $k=r+1,...,q$ whenever $r < q$. Now consider the second and third terms together. Since
\begin{align*}
a_k^{(ij)}+b_k^{(ij)}
=\left\{ \begin{array}{ll}
       \displaystyle  \sum_{l=r+1}^{q}\frac{1}{d_k^2-d_l^2}(v_{jk}v_{jl}\bh_i\t\bv_l+v_{jl}^2\bh_i\t\bv_k)
         & k=1,...,r;\\
        \displaystyle \sum_{l=}^{r}\frac{1}{d_l^2-d_k^2}(v_{jk}v_{jl}\bh_i\t\bv_l+v_{jl}^2\bh_i\t\bv_k)
         & k=r+1,...,q.\end{array} \right.
\end{align*}
it follows that the contribution from the second and the third term to the degrees of freedom equals
\begin{align*}
&\sum_{i=1}^{r_x}\sum_{j=1}^{q}\left\{\sum_{k=1}^{r}u_{ik}d_k\sum_{l=r+1}^{q}\frac{1}{d_k^2-d_l^2}(v_{jk}v_{jl}\bh_i\t\bv_l+v_{jl}^2\bh_i\t\bv_k)\right\}\\
&+\sum_{i=1}^{r_x}\sum_{j=1}^{q}\left\{\sum_{k=r+1}^{q}u_{ik}d_k\sum_{l=1}^{r}\frac{1}{d_l^2-d_k^2}(v_{jk}v_{jl}\bh_i\t\bv_l+v_{jl}^2\bh_i\t\bv_k)\right\}\\
=&\sum_{i=1}^{r_x}\left\{\sum_{k=1}^{r}u_{ik}d_k\sum_{l=r+1}^{q}\frac{1}{d_k^2-d_l^2}\sum_{j=1}^{q}(v_{jk}v_{jl}\bh_i\t\bv_l+v_{jl}^2\bh_i\t\bv_k)\right\}\\
&+\sum_{i=1}^{r_x}\left\{\sum_{k=r+1}^{q}u_{ik}d_k\sum_{l=1}^{r}\frac{1}{d_l^2-d_k^2}\sum_{j=1}^{q}(v_{jk}v_{jl}\bh_i\t\bv_l+v_{jl}^2\bh_i\t\bv_k)\right\}\\
=&\sum_{i=1}^{r_x}\left\{\sum_{k=1}^{r}\sum_{l=r+1}^{q}\frac{d_k}{d_k^2-d_l^2}u_{ik}(\bh_i\t\bv_k)
+\sum_{k=r+1}^{q}\sum_{l=1}^{r}\frac{d_k}{d_l^2-d_k^2}u_{ik}(\bh_i\t\bv_k)\right\}\\
=&\sum_{i=1}^{r_x}\left\{\sum_{k=1}^{r}\sum_{l=r+1}^{q}\frac{d_k}{d_k^2-d_l^2}u_{ik}(\bh_i\t\bv_k)
+\sum_{k=1}^{r}\sum_{l=r+1}^{q}\frac{d_l}{d_k^2-d_l^2}u_{il}(\bh_i\t\bv_l)\right\}\\
=&\sum_{k=1}^{r}\sum_{l=r+1}^{q}\left\{\frac{d_k}{d_k^2-d_l^2}\sum_{i=1}^{r_x}u_{ik}(\bh_i\t\bv_k)
+\frac{d_l}{d_k^2-d_l^2}\sum_{i=1}^{r_x}u_{il}(\bh_i\t\bv_l)\right\}\\
=&\sum_{k=1}^{r}\sum_{l=r+1}^{q}\left\{\frac{d_k}{d_k^2-d_l^2}\bu_k\t\bH\bv_k
+\frac{d_l}{d_k^2-d_l^2}\bu_l\t\bH\bv_l\right\}\\
=&\sum_{k=1}^{r}\sum_{l=r+1}^{q}\left\{\frac{d_k^2}{d_k^2-d_l^2}
+\frac{d_l^2}{d_k^2-d_l^2}\right\}\\
=&\sum_{k=1}^{r}\sum_{l=r+1}^{q}\frac{d_k^2+d_l^2}{d_k^2-d_l^2}.\\
\end{align*}
Combining the result in (\ref{eq:term1}), the proof is completed.

\subsection*{Proof of Theorem \ref{th:df2}}

Again, we assume $r_x\geq q$. When $r_x\leq q$, one can repeat the same proof using $\bH\t$. Recall that $\tilde{\bH}(\lambda)=\bU\tilde{\bD}(\lambda)\bV^{\top}$.
Consider $\partial\tilde{\bH}(\lambda)/\partial h_{ij}$ for any fixed $\lambda>0$, $1\leq i\leq r_x$ and $1\leq j\leq q$. Denote $\tilde{r}=\tilde{r}(\lambda)=\max\{k: s_k> 0.\}$. Because $\tilde{\bH}(\lambda)=\bH\sum_{k=1}^{\tilde{r}}s_k\bv_k\bv_k\t$, by the chain rule, we have
\begin{align}
\frac{\pal\tilde{\bH}(\lambda)}{\pal h_{ij}}
= &\frac{\pal\bH}{\pal h_{ij}}\sum_{k=1}^{\tilde{r}}s_k\bv_k\bv_k\t
+ \bH\sum_{k=1}^{\tilde{r}}s_k\frac{\pal\bv_k}{\pal h_{ij}}\bv_k\t
+\bH\sum_{k=1}^{\tilde{r}}s_k\bv_k\frac{\pal\bv_k\t}{\pal h_{ij}}
+\bH\sum_{k=1}^{\tilde{r}}\frac{\pal s_k}{\pal h_{ij}}\bv_k\bv_k\t \notag\\
= &\bZ^{(ij)}\bV^{(\tilde{r})}\bD^{(\tilde{r})-1}\tilde{\bD}^{(\tilde{r})}\bV^{(\tilde{r})\top}\notag\\
&-\bH\sum_{k=1}^{\tilde{r}}\left\{s_k(\bH\t\bH-d_k^2\bI)^{-}(\bH\t\bZ^{(ij)}+\bZ^{(ij)\top}\bH)\bv_k\bv_k\t\right\}\notag\\
&-\bH\sum_{k=1}^{\tilde{r}}\left\{s_k\bv_k\bv_k\t(\bH\t\bZ^{(ij)}+\bZ^{(ij)\top}\bH)(\bH\t\bH-d_k^2\bI)^{-}\right\}\notag\\
&+\bH\sum_{k=1}^{\tilde{r}}\left\{s_k'\{\frac{1}{2d_k}\bv_k\t(\bH\t\bZ^{(ij)}+\bZ^{(ij)\top}\bH)\bv_k\}\bv_k\bv_k\t\right\},\label{parij-2}
\end{align}
where $s_k'=\pal s_k/\pal d_k$. Consider the first term on the right-hand-side of (\ref{parij-2}). It can be shown that its $(i,j)$th entry equals to $\sum_{k=1}^{\tilde{r}}s_k v_{jk}^2$. Therefore, its contribution to the degrees of freedom \eqref{df} is
\begin{align}
\sum_{i=1}^{r_x}\sum_{j=1}^{q}\sum_{k=1}^{\tilde{r}}s_k v_{jk}^2=r_x\sum_{k=1}^{\tilde{r}}s_k,\label{eq:term1-2}
\end{align}
because $\sum_{j=1}^{q}v_{jk}^2=1$. Similar to the proof of Theorem \ref{th:df1}, the $(i,j)$th entry of the second and third terms on the right-hand-side of \eqref{parij-2} can be shown to be
\begin{align}
\bu_i\t\bD(\tilde{\ba}^{(ij)}+\tilde{\bb}^{(ij)})
\end{align}
where $\tilde{\ba}^{(ij)}\in \mathbb{R}^q$, $\tilde{\bb}^{(ij)}\in \mathbb{R}^q$, and
\begin{align*}
\tilde{a}_k^{(ij)}+\tilde{b}_k^{(ij)}
=\left\{ \begin{array}{ll}
     \displaystyle   \sum_{l\neq k}^{q}\frac{s_k-s_l}{d_k^2-d_l^2}(v_{jk}v_{jl}\bh_i\t\bv_l+v_{jl}^2\bh_i\t\bv_k)
         & k=1,...,\tilde{r};\\
      \displaystyle   \sum_{l=1}^{\tilde{r}}\frac{s_l}{d_l^2-d_k^2}(v_{jk}v_{jl}\bh_i\t\bv_l+v_{jl}^2\bh_i\t\bv_k)
         & k=\tilde{r}+1,...,q.\end{array} \right.
\end{align*}
After some algebra, it follows that the contribution from the second and the third term to the degrees of freedom equals \\
\begin{align}
&\sum_{i=1}^{r_x}\sum_{j=1}^{q}\left\{\sum_{k=1}^{\tilde{r}}u_{ik}d_k  \sum_{l\neq k}^{q}\frac{s_k-s_l}{d_k^2-d_l^2}(v_{jk}v_{jl}\bh_i\t\bv_l+v_{jl}^2\bh_i\t\bv_k)\right\}\notag\\
&+\sum_{i=1}^{r_x}\sum_{j=1}^{q}\left\{\sum_{k=\tilde{r}+1}^{q}u_{ik}d_k \sum_{l=1}^{\tilde{r}}\frac{s_l}{d_l^2-d_k^2}(v_{jk}v_{jl}\bh_i\t\bv_l+v_{jl}^2\bh_i\t\bv_k)\right\}\notag\\
=&\sum_{k=1}^{\tilde{r}}\sum_{s=\tilde{r}+1}^{q}\left\{\frac{d_k^2(s_k-s_l)+d_l^2s_k}{d_k^2-d_l^2}\right\}
+\sum_{k=1}^{\tilde{r}}\sum_{l\neq k}^{\tilde{r}}\left\{\frac{d_k^2(s_k-s_l)}{d_k^2-d_l^2}\right\}.\label{eq:term23-2}
\end{align}

\noindent Consider the fourth term on the right-hand-side of \eqref{parij-2}. Note that
$$
\bv_k\t(\bH\t\bZ^{(ij)}+\bZ^{(ij)\top}\bH)\bv_k=2v_{jk}(\bv_k\t\bh_i).
$$
The $(i,j)$th entry of the fourth term is given by
$$
\sum_{k=1}^{\tilde{r}}s_k' u_{ik} v_{jk}^2(\bv_k\t\bh_i).
$$
It then follows that the contribution of the fourth term to the degrees of freedom equals
\begin{align*}
& \sum_{i=1}^{r_x}\sum_{j=1}^{q}\sum_{k=1}^{\tilde{r}}s_k'u_{ik}v_{jk}^2(\bv_k\t\bh_i)\\
=& \sum_{i=1}^{r_x}\sum_{k=1}^{\tilde{r}}s_k'u_{ik}(\bv_k\t\bh_i)\\
=& \sum_{k=1}^{\tilde{r}}s_k'\sum_{i=1}^{r_x}u_{ik}\bh_i\t\bv_k\\
=& \sum_{k=1}^{\tilde{r}}d_k s_k'.
\end{align*}
Combining with the results in (\ref{eq:term1-2}) and (\ref{eq:term23-2}), the proof is completed.

\end{document}